\documentclass[12pt,preprint,usenatbib]{emulateapj}

\usepackage{amssymb}
\usepackage{times}
\usepackage{pifont} 
\usepackage{rotating}

\newcommand{\teff}{$T_{\mathrm{eff}}$}
\newcommand{\muhz}{$\mu$Hz}
\newcommand{\numax}{$\nu_{\mathrm{max}}$}
\newcommand{\dnu}{$\Delta\nu$}

\newcommand{\msol}{M$_\odot$}
\newcommand{\rsol}{R$_\odot$}

\newcommand{\kepler}{\textit{Kepler}}

\shorttitle{Revisiting old friends}
\shortauthors{Stello et al.}

\begin{document}

%\title{Revisiting old friends: unambiguous detection of oscillations in red giants of the open cluster M67}%Victor
%\title{Revisiting old friends: oscillating red giants in the open cluster M67 observed with the K2 mission}
%\title{Revisiting old friends with K2: oscillations in the red giants of the open cluster M67}
%\title{Revisiting old friends from M67 with K2: oscillations in the open cluster red giants}
%\title{Revisiting old friends from M67 with K2 reveals oscillating red giants in `solar-like' open cluster}
\title{The K2 M67 Study: Revisiting old friends with K2 reveals oscillating red giants in the open cluster M67}

\author{
Dennis~Stello\altaffilmark{1,2,3}, 
Andrew~Vanderburg\altaffilmark{4},    %LC generation
Luca~Casagrande\altaffilmark{5},      %Teff derivation
Ron~Gilliland\altaffilmark{6},        %Target selection
Victor~Silva~Aguirre\altaffilmark{3}, %Gridmodeling
Eric~Sandquist\altaffilmark{7},       %very helpful comments
Emily~Leiner\altaffilmark{8},         %Masterlist generation
Robert~Mathieu\altaffilmark{8},        %Target selection
David~R.~Soderblom\altaffilmark{9}    %comments on manus
%Hans~Kjeldsen\altaffilmark{1,2},      %discussion
%Tim~R.~Bedding\altaffilmark{1,2},     %Said no.
%Jie\altaffilmark{1,2},                %Said no.
%Daniel~Huber\altaffilmark{1,3,2},     %Said no.
%Marc~Pinsonneault\altaffilmark{4},    %?
%Sarbani~Basu\altaffilmark{7},         $SC target selection
}
\altaffiltext{1}{Sydney Institute for Astronomy (SIfA), School of Physics, University of Sydney, NSW 2006, Australia}
\altaffiltext{2}{School of Physics, University of New South Wales, NSW 2052, Australia}
\altaffiltext{3}{Stellar Astrophysics Centre, Department of Physics and Astronomy, Aarhus University, DK-8000 Aarhus C, Denmark}
\altaffiltext{4}{Harvard–Smithsonian Center for Astrophysics, Cambridge, MA 02138, USA}
\altaffiltext{5}{Research School of Astronomy \& Astrophysics, Mount Stromlo Observatory, The Australian National University, ACT 2611, Australia}
\altaffiltext{6}{Department of Astronomy and Astrophysics, The Pennsylvania State University, University Park, PA 16802, USA}
\altaffiltext{7}{San Diego State University, Department of Astronomy, San Diego, CA 92182, USA}
\altaffiltext{8}{Department of Astronomy, University of Wisconsin-Madison, Madison, Wisconsin 53706, USA}
\altaffiltext{9}{Space Telescope Science Institute, Baltimore, Maryland 21218, USA}
%\altaffiltext{4}{Department of Astronomy, The Ohio State University, Columbus, OH 43210, USA} ;Marc
%\altaffiltext{7}{Department of Astronomy, Yale University, P.O. Box 208101, New Haven, CT 06520-8101} ;Sarbani
%\clearpage

\begin{abstract}
Observations of stellar clusters have had a tremendous impact in forming our
understanding of stellar evolution.  The open cluster M67 has a
particularly important role as a calibration %anchor point 
benchmark 
for stellar evolution theory due to its near solar composition and age. 
%making it the birth place of solar twins.  
As a result, it has been observed
extensively, including attempts to detect solar-like oscillations in its
main sequence and red giant stars.  However, any asteroseismic inference has so
far remained elusive due to the difficulty in measuring these extremely
low amplitude oscillations.  
Here we report the first unambiguous detection of solar-like oscillations
in the red giants of M67.  We use data from the \kepler\ ecliptic mission,
K2, to measure the global asteroseismic properties.
We find a model-independent seismic-informed distance of 
816$\pm11\,$pc, or $(m-M)_0=9.57\pm 0.03\,$mag, an average red-giant
mass of $1.36\pm0.01\,$\msol, in agreement with the dynamical mass
from an eclipsing binary near the cluster turn-off, 
%and we find an age of $3.50\pm0.13\,$Gyr.  
%but we find a lower-than-usual age.
and ages of individual stars compatible with isochrone fitting.
We see no evidence of strong mass loss on the red
giant branch.  
We also determine seismic $\log g$ of all the cluster giants with a typical
precision of $\sim 0.01\,$dex.
Our results generally show good agreement with independent methods and
support the use of seismic scaling relations to determine global
properties of red giant stars with near solar metallicity. 
We further illustrate that the data are of such high quality, that
future work on individual mode frequencies should be possible, which would 
extend the scope of seismic analysis of this cluster.
\end{abstract}


%\clearpage

\section{Introduction} 
M67 is an open cluster with 
approximately solar age and metallicity, 
%an age and metallicity similar to the Sun, 
making it a prime target in stellar astrophysics for decades. 
After the demonstrated success of applying seismic techniques to the Sun
\citep[e.g.][]{Duvall84,Dalsgaard85}, recent decades also saw studies aimed to detect
solar-like oscillations in these cluster stars, mainly around the main
sequence and the turn-off \citep{Gilliland91,Gilliland93}. However,
success was limited, due to the extremely low amplitude oscillations.
In the hope of obtaining unambiguous detections a 6-week 10-telescope 
multi-site campaign was launched, aimed to detect oscillations in the
cluster's giant stars \citep{Stello06,Stello07}.  With only marginal
detections at best, this campaign concluded the past two decades of
ground-based attempts. %in realization that even K giants would require
%ultra-high precision photometry from space or dedicated ground-based radial  
%velocity measurements to generate truly useful results.
Fortunately, the \kepler\ space telescope turned out to be an incredible source
of asteroseismic data, %also of open cluster stars, 
%showing clear detection of
clearly showing 
oscillations in open cluster red giants. This allowed inferences to be made on 
cluster age, mass loss along the giant branch, and seismic membership
\citep[e.g.][]{Stello10,Basu11,Miglio12,Stello11a}; including results showing that the
oscillation amplitudes anticipated for the giants by previous
ground-based campaigns were generally overestimated \citep{Stello11b}.
With \kepler's ecliptic second-life mission, K2 \citep{Howell14},
its potential for seismic
studies of open clusters increased because of the many clusters within its
new viewing zone.  
In particular, data from its observing campaign 5 has been much anticipated
because it included M67.  

In this paper, we report the first results in a series arising from a
large collaboration aimed at observing and studying M67 using K2 time
series photometry.  After a description of our general target selection,
we focus on the analysis of the red giant cluster members.  Initially we
make comparisons with previous attempts to detect oscillations, and follow
on with measuring the global asteroseismic properties from which we
determine stellar radius, mass, and age.  We compare these results with
independent literature values and investigate the asteroseismic scaling
relations, widely used for radius and mass estimation.

%In this paper, we report the time series analysis of the red giants
%in M67 observed by K2. After an initial comparison with previous %seismic detection
%attempts we measure the global asteroseismic properties from which we
%determine stellar radius, mass, and age.  We 
%compare our results with independent literature values and 
%investigate the asteroseismic scaling relations, widely used for
%radius and mass estimation.

\section{Observations and light curve preparation}\label{observations}
The goal of the general target selection for the K2 M67 study was to include all stars
for which extensive kinematic information \citep{Geller15}
supports even a modest probability of cluster membership.  Sample completeness in the majority of the
cluster has been assured by using a 400 by 400 pixel (26.5' by 26.5'
square)\footnote{The superaperture size corresponds to approximately six
  core radii of the cluster \citep[see][~and references therein]{Geller15}}
superaperture centered on the cluster, which is recorded and downloaded to
ground in its entirety.  Outside the dedicated aperture, all known or suspected
cluster members were added as single targets; we adopted
$\langle P_\mathrm{RV},P_\mathrm{PM}\rangle >20$\% to be inclusive. %like any other 
%individual field star.  
The superaperture and all added 
targets for M67 were observed for about 75 days (Campaign 5, 27 April - 10
July, 2015) in the spacecraft's long-cadence
mode ($29.4\,$min), while a few were also observed in short cadence
($58.85\,$sec).  %as part of a program
%to detect oscillations in the main sequence turn-off and subgiant stars.
Extensive cross-checks verified that all the giant members
were included in our proposal (e.g. all \citealt{Stello06} targets).

For this investigation, the extraction of photometric time series (light curves) was performed
using the technique by \citet{VanderburgJohnson14} with the updates
described in \citet{Vanderburg16} both for single targets and those
captured by the superaperture. For some of the giants, we processed the
light curves in a slightly different manner. The
\citet{VanderburgJohnson14} method accounts for low-frequency variations in
K2 light curves by modeling them with a basis-spline, usually with
breakpoints every 1.5 days. For giants with oscillation frequencies around
$\sim$ 10 \muhz, we used a faster spline with breakpoints every 0.3 days to
model the low-frequency variations, which in these cases are dominated by
the seismic oscillations. Modeling the oscillations with a faster spline
decreased the noise level in the light curves. Of the resulting light
curves, about 70\% arose from the superaperture.

The post processing of the data follows that described in
\citet{Stello15}. In short, we apply a high-pass filter and fill 
short gaps, %mostly caused by the regular thruster firings, 
using
linear interpolation.  The applied filter had a characteristic cut-off
frequency of $\sim3$\muhz\ for most stars, and $\sim6$\muhz\ 
for the low-luminosity red giant branch (RGB) stars ($V>12.4$). The light
curves of the two most luminous stars were not high-pass filtered, to avoid
affecting the oscillation signal.

\section{Stellar sample and comparison with previous results}
\begin{figure}
\includegraphics[width=8.8cm]{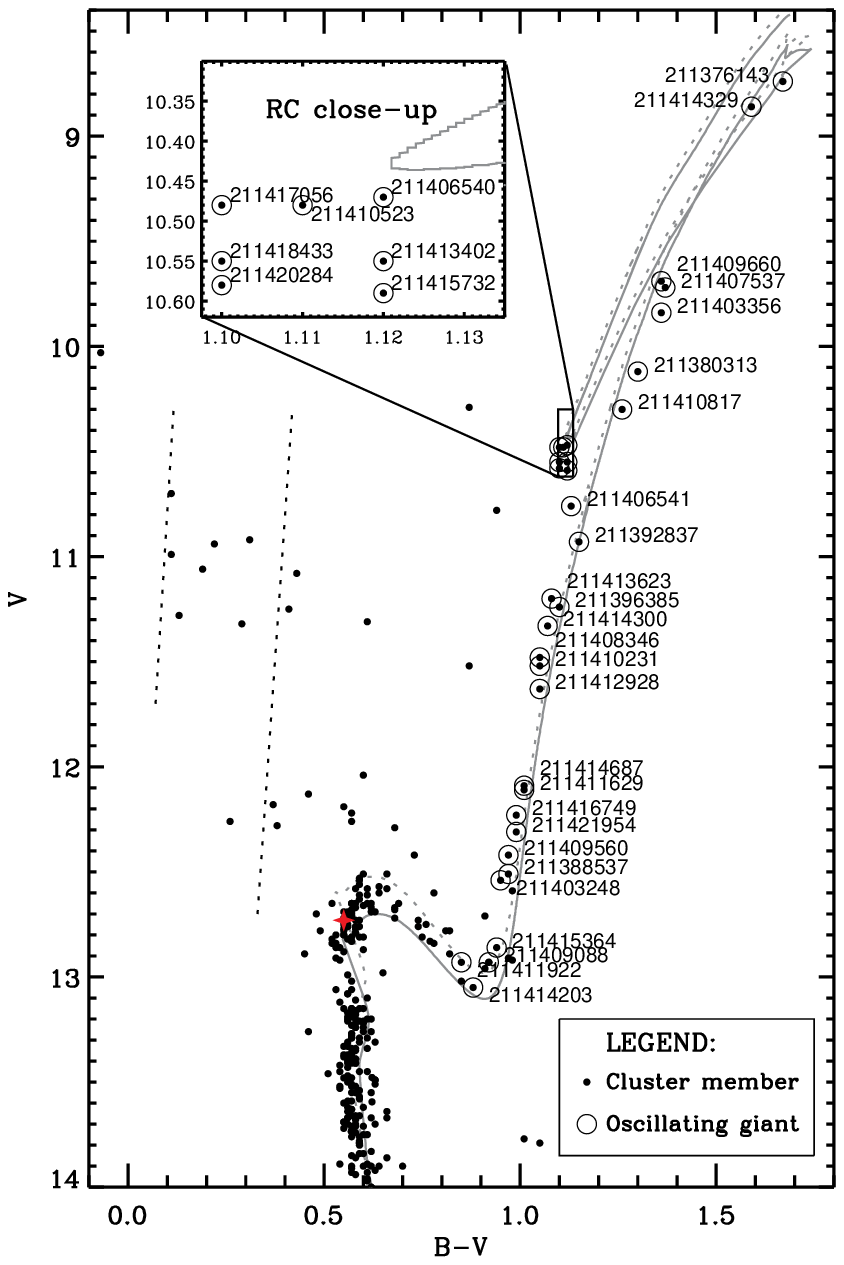}
\caption{Color-magnitude diagram of M67 cluster members. Photometry (not
  corrected for reddening) is from \citet{Montgomery93} and membership is from \citet{Geller15}.
  Stars with detected oscillations are encircled and
  indicated by their EPIC-ID.  %Those shown in Figure 2 are filled red.
  Gray curves show %a standard solar-scaled 
  BaSTI isochrones of $3.5\,$Gyr (dotted) and
  $4.0\,$Gyr (solid)
  %and no mass loss 
  \citep{Pietrinferni04}, shifted 9.7
  mag vertically and 0.03 mag horizontally. The instability strip is
  shown by dotted black lines \citep{RodriguezBreger01}. The eclipsing
  binary HV Cnc is marked by a red star symbol.
%MAYBE INDICATE SC TARGETS
\label{cmd}} 
\end{figure}

For the following asteroseismic analysis, we selected all giant stars
with available K2 data if they were listed as kinematic members
by \citet{Geller15} and were following the giant branch including the
helium-burning phase in the Hertzsprung-Russell diagram 
%(illustrated by the color-magnitude diagram in Figure~\ref{cmd}).
(Figure~\ref{cmd}).
%All our giant targets have \teff\ $<5200\,$K and $\log\,g<3.8$. 
This seismic sample is unique in the way it spans the entire RGB, %from
bottom-to-tip, and the helium-core burning `red clump' (RC) phase of a
simple stellar population.    
Whether it includes any asymptotic giant branch stars
is unknown. However, generally stars brighter than
the RC but fainter than the RGB tip are most likely RGB stars.
We do note that while EPIC211409660 and 211407537 have effective
temperatures compatible with the isochrone RGB they have lumonisities
near the asymptotic giant branch luminosity bump
\citep{SalarisCassisi05}, located at $V\sim9.6-9.7$. %for the isochrone in Figure~\ref{cmd}. 
%{\bf They could therefore be ceasing core helium burning and
%transitioning into shell helium burning asymptotic giant branch stars.}
We calculated the %Fourier transforms %(using IDL lnp_test)
%of the light curves to derive 
power density frequency spectra for all the giants, to reveal the
presence of oscillations. 
The color-magnitude diagram shows all giant members %from \citet{Geller15} brighter than $V=14$ 
showing oscillations (Figure~\ref{cmd}, black dots with 
open circles).  
%To guide the eye we show the instability strip \citep{RodriguezBreger01}
%and a BaSTI 4-Gyr isochrone \citep{Pietrinferni04}, shifted by-eye to
%roughly match the photometry of the cluster.
%[IS THIS TOO DISTRACTIVE/NOT RELEVANT OR DO WE WNAT TO ENCURANGE PEOPLE TO
%DIG INTO THE DATA?] 
%Less than five of all the plotted members have not been observed by K2.

\begin{figure*}
\includegraphics[width=17.6cm]{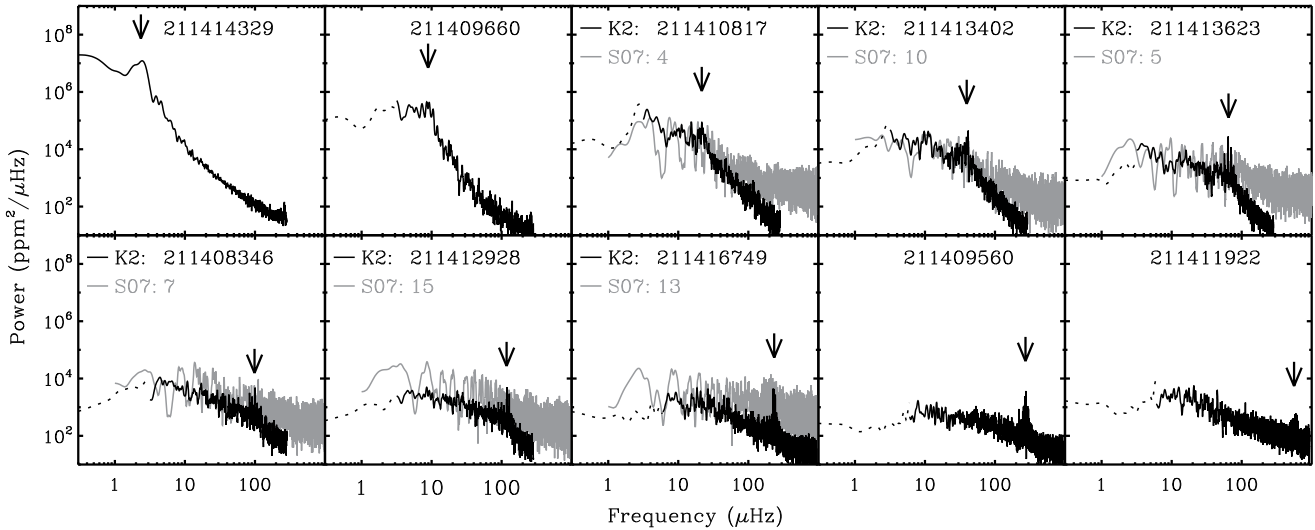}
\caption{Power spectra from K2 data of 10 representative giants
  (black).  The region affected by high-pass filtering is
  indicated (dotted curve).  Ground-based results \citep[][ S07]{Stello07}
  are also shown (gray).  The K2 and ground-based spectra are smoothed to a
  common frequency resolution. Arrows indicate the oscillation power. 
  Both EPIC and S07 IDs are shown.
\label{log-log-spectra}} 
\end{figure*}

The power spectra of a representative sample of our giant targets are shown in
Figure~\ref{log-log-spectra} (black curves) with
ordinate and abscissa ranges identical for all stars.
The panels are ordered by brightness, indicative of relative luminosity for
cluster stars, going from brightest at top left to faintest at bottom right.
It is unambiguous that we see oscillations for the entire
range of evolutionary stages spanned by the cluster giants.
The figure illustrates nicely 
%beautifully 
how amplitude and timescale of both the
granulation (downward sloping background noise) and of the oscillations
(hump indicated by arrows) scale with luminosity
\citep{KjeldsenBedding11,Stello11b,Mathur11,Kallinger14}. It is
also evident at high frequencies that the largely photon-dominated 
(white) noise increases towards fainter stars.  

%COMPARISON WITH STELLO07/GILLILAND-G12
Our clear detection of oscillations naturally leads to the
question of whether previous ground-based campaigns did indeed detect this signal.
For this, we focus on \citet{Stello07} who specifically
targeted the giants.
In Figure~\ref{log-log-spectra} we show 
stars in common with \citet{Stello07} (gray).  %Both
%K2 and ground-based data sets are smoothed to a common frequency resolution
%to facilitate a direct like-for-like comparison. %assesment of how close the ground-based campaign was
%to detecting the oscillations.  
We conclude that the reported excess power by \citet{Stello07}
could quite plausibly be oscillations for the most luminous stars
in their sample, near the red clump luminosity, while it seems unlikely oscillations
were detected for the lower luminosity RGB stars. 
A similar comparison with the one giant (211408346), which fell
serendipitously in the turn-off star sample
studied by \citet{Gilliland93}, showed %that the
noise levels %were
5-10 times too high in the ground-based data 
to plausibly see evidence of the oscillations.

\section{Extracting seismic observables}
\begin{figure}
\includegraphics[width=8.8cm]{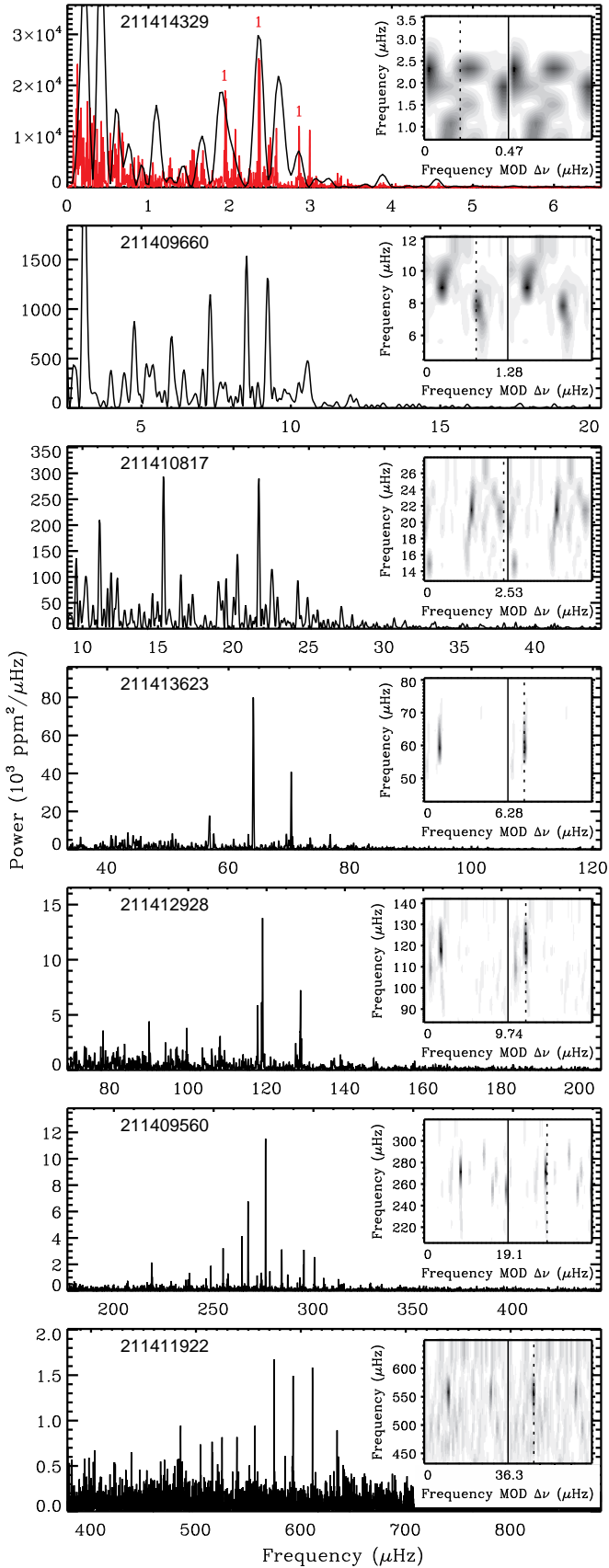}
\caption{Power spectra and \'echelle diagrams (insets) of seven giants
  (also shown in Figure~\ref{log-log-spectra}).  %Only regions dominated 
%  by oscillation frequencies are shown (\numax $\pm$ 4\dnu). 
  The top panel also shows (red) a star observed by
  \kepler\ with its dipole modes indicated following \citet{Stello14}.
  The \'echelle diagrams stack consecutive \dnu -wide bins of the spectrum
  and are plotted twice side-by-side, as indicated by the black
  vertical line. The vertical dotted line indicates the approximate
  location of the radial mode ridge. 
\label{lin-lin-spectra}} 
\end{figure}

To demonstrate %the quality of the K2 data and 
that we can perform
meaningful seismic analyzes including extracting the large frequency
separation, \dnu, and the frequency of maximum power,
\numax\ (and in many cases possibly individual frequencies), we show regions of power
spectra and \'echelle diagrams centered around \numax\ in
Figure~\ref{lin-lin-spectra}.  
%For stars with \numax\ $\lessapprox 10\,$\muhz, we expect the uncertainty in \dnu\ to
For stars with \numax\ $\lesssim 10\,$\muhz, we expect the uncertainty in \dnu\ to
be relatively large, limited by the frequency resolution of the K2
data \citep{Stello15}.  For the most luminous star (top) we therefore over plot a
comparison spectrum in red of a similar star observed for four years
by \kepler.  
%When fully resolved, the spectrum of such a luminous 
%M giant looks like sets of triplets for each set of
%dipole-quadrupole-radial modes \citep[see][]{Stello14}. 
With a formal frequency
resolution of $\sim 0.15\,$\muhz\ from the K2 data we only obtain the
broad features of the underlying mode structure, `just' enough to measure
\dnu, which also explains the somewhat blurry \'echelle diagram for this star.
Comparing the \'echelle diagrams from top (near the RGB tip) to bottom
(near RGB bottom) we see the known gradual increase in the
location of the radial mode ridge (see vertical dotted lines), also known as
the offset, $\varepsilon$, in the asymptotic relation for acoustic
oscillations as observed by e.g.
\citet{Huber10,Mosser10universal,Corsaro12}.  For the star in the  
bottom panel we start to see the slight decrease in $\varepsilon$, as
expected from models \citep{White11}. 
%POSSIBLY REMOVE TO SAVE SPACE
 %We also note from the \'echelle diagrams that some of the stars
 %seem to show at least some suppression of dipole and quadrupole modes
 %\citep{Mosser12a,Stello16b}, which is evidence of strong internal
 %magnetic fields \citep{Fuller15} generated by a core dynamo when the stars
 %had convective cores during their main sequence phase \citep{Stello16a}.  
 
\begin{figure*}
\includegraphics[width=17.6cm]{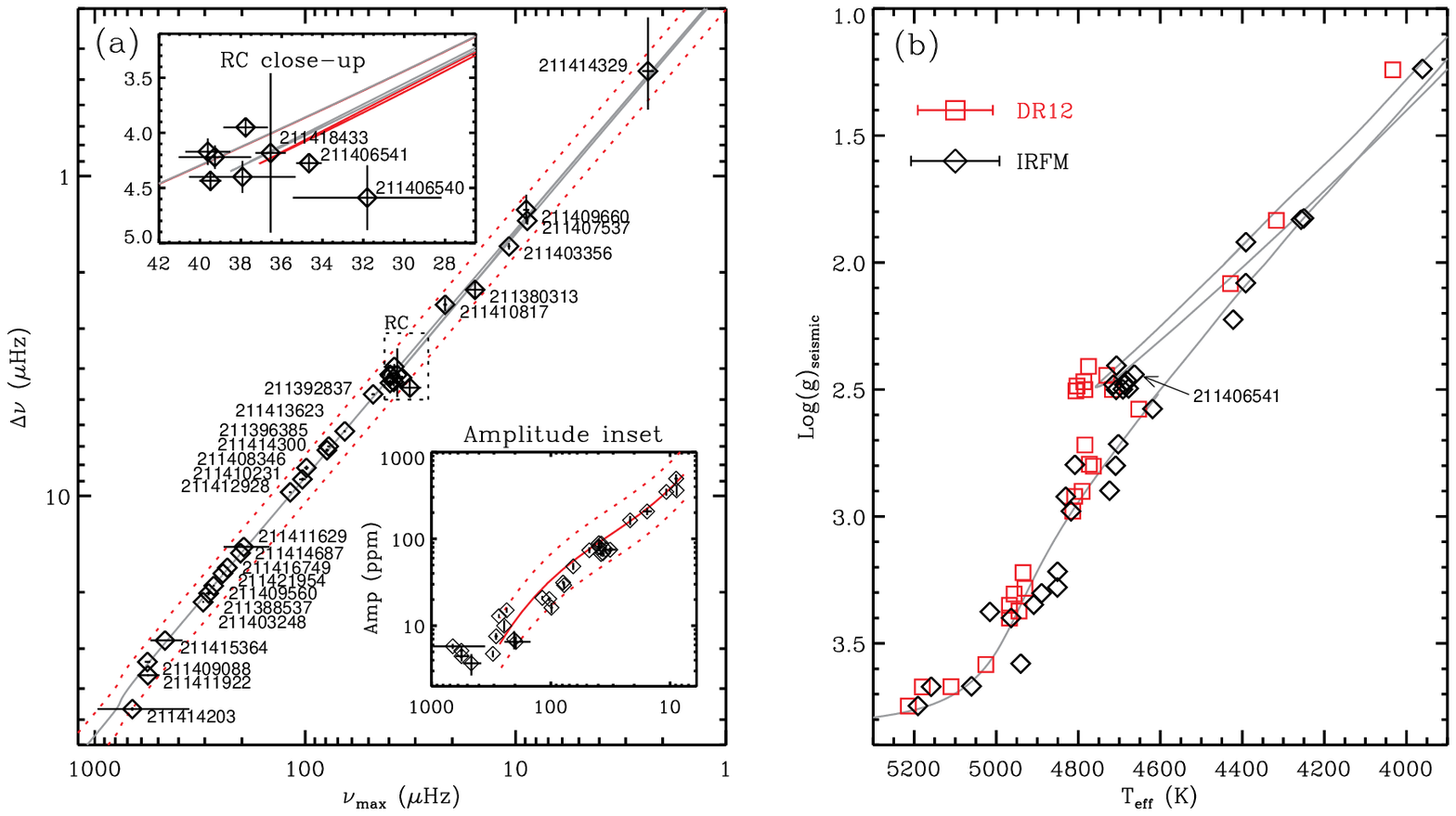}
\caption{(a): Observed \dnu\ versus \numax.  The solid isochrone from
  Figure~\ref{cmd} is shown (gray), where \numax\ and \dnu\ are derived
  using the scaling relations corrected with the \citet{Sharma16}
  prescription for \dnu. 
  The red dotted curves show typical extremes of the spread for a large sample of giants
  observed during K2 campaign-1 analyzed the same way as the M67
  giants (Stello et al. in prep), and correspond to an uncorrected scaling
  mass range of 0.75-2.9\msol (upper-to-lower curves).
  EPIC IDs are shown for reference. $1\sigma$
  error bars are mostly much smaller than the symbols.
  The RC close-up also shows the isochrone with mass loss of $\eta=0.2$ (red).
  In the lower inset the abscissa is replaced by oscillation amplitude per
  radial mode in parts per million.  The red curve shows the average
  (solid) trend and typical extremes (dotted) from the K2 campaign-1 data.
  (b) $\log g_\mathrm{seismic}$-\teff\ diagram showing all
  IRFM- (this paper) and SDSS-DR12-based \teff\ values. Average  $1\sigma$
  error bars are shown.
\label{kiel_dnu-numax_amp-numax}} 
\end{figure*}

Using the method by \citet{Huber09} to analyze the power spectra,
we were able to detect \dnu, \numax, and oscillation amplitude for all the
observed giants of M67, except four stars near the bottom of the RGB, which
were only observed in
the spacecraft's long-cadence mode. Non-detections among these stars are
expected because they oscillate beyond the long-cadence Nyquist
(half the sampling) frequency of $\sim 283\,$\muhz, making it
harder to measure the seismic signal.  We were, however, able to measure the
signal for another two such `super-Nyquist' stars \citep[see][~for a similar
technique applied to field giants]{Yu16}. 
%\citep[see also~][~for similar technique]{Murphy13,Chaplin14,Yu16}. 
%This difficulty arises due to both increasing  
%granualation noise reflected off the Nyquist frequency from lower frequency
%and decreasing signal caused by the averaging of the oscillation amplitude
%by the $\sim30\,$min long integration time.
In addition, the super-Nyquist issue was mitigated for four giants, for
which K2 short-cadence data were also available.   
Despite that, the star sitting at the very bottom of
the RGB (211414203), showed only marginal detection due to the intrinsically lower
oscillation amplitude and increasing photon noise towards less evolved and
fainter stars. 
% While we would caution the use of its seismic parameters we have included
% it for completeness and would encourage other methods to extract the signal.
Also, the largest and most luminous giant in our sample (211376143)
oscillates at such low frequency that \dnu\ cannot be measured.
%from the roughly 75 days of K2 data. %, but the dominating periodicity (hence
%\numax) is clearly visible in the time series and its Fourier transform.
We list our measured \numax\ and \dnu\ values in Table~\ref{tab1} (columns 4-5),
and plot them together with oscillation amplitude in
Figure~\ref{kiel_dnu-numax_amp-numax}a. %(the amplitude data are available
%in the online version).  
%THIS COULD GO IF NEED SPACE
 %While 211411922 appears to be a late subgiant in
 %Figure~\ref{cmd}, its seismology demonstrates it is an RGB star, with lower
 %\numax\ (hence $\log\,g$) than the least evolved giant in our sample;
 %this could indicate a larger than expected error on its $B-V$ color from
 %\citet{Geller15}. 
It is evident that the stars with small error bars line up almost
perfectly on a straight line in \dnu-\numax\ space as expected for stars
with almost the same mass \citep[e.g.][]{Stello09,Huber10}.  
The isochrone in Figure~\ref{kiel_dnu-numax_amp-numax}a (gray) has a 
%EITHER
%an empirical average trend from field stars covering a range of masses, 
%OR 
mass-change along the RGB of less than 0.03\msol. Significant differences
in mass move stars above or below a straight line in this diagram \citep[e.g.][]{Stello09,Hekker10a}. 
This demonstrates that these stars are excellent candidates for testing the
seismic scaling relations for different evolution stages where most
properties are otherwise similar for all the stars. 
To further support the detection of oscillations, we show that the
amplitude-\numax\ trend follows that of solar-like oscillations in field
red giants (Figure~\ref{kiel_dnu-numax_amp-numax}a; lower inset).
We were not able to obtain robust measurements of oscillation amplitude for
the two most luminous stars.

\section{Benchmarking seismic scaling relations and mass loss}
Having several giants in various evolution stages enables us to estimate the
stellar masses along the RGB and RC, and in combination with
eclipsing binaries, to benchmark the seismic scaling relations
\citep{Jeffries13,Sandquist13,Brogaard12}. In addition, one can look for
evidence of mass loss \citep{Miglio12}.  
To do this we first need \teff\ for as many stars as possible.

\subsection{Temperature scales}
We used two sources for \teff.  
%Our main source came from applying directly the Infrared Flux Method (IRFM,
%using the same methodology described in Casagrande et al. 2010)  on 31
%stars in our seismic sample for which optical (Tycho2 and/or APASS) plus
%2MASS photometry was available 
Our main source came from applying the Infrared Flux Method (IRFM) methodology
by \citet{Casagrande10} on 31 stars in our 
seismic sample for which optical (Tycho2 and/or APASS) plus 2MASS photometry was available.
%\citep{Cutri03}.  
Here we assumed [Fe/H$]=0$, E$(B-V)=0.03$
\citep[e.g.][~and references therein]{CasagrandeVandenBerg14}, and
seismic $\log g$ obtained from \numax\ and an initial \teff. The
method is only mildly dependent on the adopted gravity and metallicity, 
and convergence in \teff\ was reached after one iteration
with seismic gravities. The adopted \teff\ and uncertainties were derived by averaging the
results obtained by implementing the aforementioned photometric systems into
the IRFM following \citet{Casagrande14SAGA}. The scatter was used to
estimate the uncertainties after increasing it by $20\,$K to account for
the systematic uncertainty on our \teff\ zero-point %On average the uncertainty
%is $108\,$K. 
(Table~\ref{tab1} column 6).  
For comparison we also adopted spectroscopically-determined
\teff\ values from SDSS-DR12
\citep[][ \textsc{fparam}]{Alam15}, for the 27-star subset for which
  \teff\ was available from both sources %, with quoted
%uncertainties of $91\,$K 
(Figure~\ref{kiel_dnu-numax_amp-numax}b).
%We adopted the so-called `corrected' values, which are
%$87\,$K hotter than the raw values.    
%Comparing results from the two sources revealed the IRFM-based \teff\ is
%$87\pm25\,$ K cooler on average, hence similar to the SDSS DR12 raw
%values, but with a slight trend with $\log g$
%(Figure~\ref{kiel_dnu-numax_amp-numax}b). The IRFM results 
%show slightly larger scatter relative to the adopted BaSTI isochrone
%(gray curve).  The systematic \teff\ difference
%between IRFM and SDSS translate into small differences in the
%seismically-inferred stellar properties (roughly 1\% in radius, 2-3\% in
%mass, and below 0.005 dex in $\log g$).

\begin{table*}
{\footnotesize
\begin{center}
\caption{Seismic properties of M67 red giants. Subscripts `sc', `corr', and
  `grid' indicate the seimic methods, scaling, corrected scaling, and grid-based
  modelling, respectively. \label{tab1}}
\begin{tabular}{rccccc|ccc|cc|cccc}
\tableline\tableline
\multicolumn{1}{c}{EPIC}    & WOCS    & Class  & \numax  & \dnu    & \teff/K & $R_\mathrm{sc}$ & $M_\mathrm{sc}$ & $\log g_\mathrm{sc}$ & $R_\mathrm{corr}$ & $M_\mathrm{corr}$ & $R_\mathrm{grid}$ & $M_\mathrm{grid}$ & $\log g_\mathrm{grid}$ & Age$_\mathrm{grid}$  \\
\multicolumn{1}{c}{ID}      & ID      &        & (\muhz) & (\muhz) & (IRFM)  & (\rsol)       & (\msol)        & (cgs)              & (\rsol)         & (\msol)          & (\rsol)         & (\msol)        & (cgs)                & (Gyr)               \\
\multicolumn{1}{c}{[1]$^a$} & [2]$^b$ & [3]$^c$ & [4]     & [5]     & [6]     & [7]           & [8]            &  [9]               & [10]            & [11]             & [12]            & [13]           & [14]                 & [15]                \\
\tableline		  	  	    		                       
               211376143   &   1075  &   SM    &   0.81(10) &           &           &           &             &                    &                 &                  &                  &                &                      &                     \\   
               211414329   &   1036  &   SM    &   2.35(15) &  0.47(15) & 3960(92)  & 52.0(3.4) & 1.7(2.2)    & 1.237(32)          & 50.3(3.3)       &  1.6(2.0)        &  44.7(5.0)       & 1.25(28)       & 1.235(28)            &  5.3(4.2)           \\  
               211407537   &   1008  &   SM    &   8.81(11) &  1.38(4)  & 4250(126) & 23.43(45) & 1.34(17)    & 1.826(49)          & 22.02(43)       &  1.19(15)        &  21.7(1.3)       & 1.15(14)       & 1.825(6)             &  7.2(3.2)           \\  
               211409660   &   1005  &   BM    &   8.90(29) &  1.28(13) & 4256(116) & 27.73(98) & 1.90(81)    & 1.831(45)          & 26.58(94)       &  1.75(73)        &  23.0(2.5)       & 1.30(29)       & 1.828(14)            &  4.6(4.0)           \\  
               211403356   &   1045  &   SM    &  10.75(15) &  1.66(4)  & 4391(97)  & 20.12(36) & 1.23(13)    & 1.920(40)          & 18.85(34)       &  1.08(11)        &  18.80(74)       & 1.06(9)        & 1.915(7)             &  9.6(3.0)           \\  
               211380313   &   1065  &   SM    &  15.6(1.3) &  2.27(15) & 4391(78)  & 15.6(1.4) & 1.07(40)    & 2.081(36)          & 14.5(1.3)       &  0.93(34)        &  15.9(1.2)       & 1.14(19)       & 2.095(28)            &  7.5(4.1)           \\  
               211410817   &   2004  &   SM    &  21.62(44) &  2.53(12) & 4422(92)  & 17.53(40) & 1.88(38)    & 2.225(40)          & 16.73(38)       &  1.71(34)        &  15.4(1.4)       & 1.45(27)       & 2.224(9)             &  3.1(2.1)           \\  
             **211406540   &   1029  &   SM    &  31.8(3.6) &  4.59(29) & 4707(107) &  8.05(93) & 0.60(26)    & 2.406(61)          &  8.14(94)       &  0.62(26)        &  10.98(25)       & 1.23(15)       & 2.441(38)            &  5.7(2.7)           \\  
               211406541   &   2014  &   BM    &  34.66(62) &  4.27(7)  & 4663(68)  & 10.07(19) & 1.02(9)     & 2.441(35)          &  9.62(19)       &  0.93(8)         &   9.88(19)       & 0.99(4)        & 2.444(6)             & 12.3(1.8)           \\  
             **211418433   &   1022  &   SM    &  36.54(74) &  4.18(72) & 4681(73)  & 11.11(24) & 1.31(91)    & 2.465(38)          & 11.16(24)       &  1.33(92)        &  11.12(22)       & 1.33(6)        & 2.466(9)             &  4.4(7)             \\  
             **211410523   &   1003  &   SM    &  37.8(1.1) &  3.95(8)  & 4714(117) & 12.92(40) & 1.84(23)    & 2.481(62)          & 13.02(41)       &  1.87(23)        &  12.31(73)       & 1.63(23)       & 2.468(15)            &  2.3(1.1)           \\  
             **211415732   &   1009  &   SM    &  37.9(2.6) &  4.40(14) & 4687(110) & 10.42(73) & 1.20(30)    & 2.481(59)          & 10.46(73)       &  1.21(30)        &  11.09(16)       & 1.39(8)        & 2.487(20)            &  3.7(8)             \\  
             **211420284   &   2019  &   SM    &  39.3(1.8) &  4.22(11) & 4677(89)  & 11.72(54) & 1.57(27)    & 2.496(46)          & 11.80(54)       &  1.59(27)        &  11.17(21)       & 1.39(8)        & 2.482(15)            &  3.7(8)             \\  
             **211413402   &   2005  &   SM    &  39.49(54) &  4.44(9)  & 4691(153) & 10.68(23) & 1.31(14)    & 2.499(79)          & 10.73(23)       &  1.33(14)        &  11.10(13)       & 1.43(4)        & 2.503(6)             &  3.3(4)             \\  
             **211417056   &   2010  &   SM    &  39.6(1.1) &  4.17(12) & 4707(113) & 12.14(37) & 1.70(25)    & 2.501(59)          & 12.23(37)       &  1.73(25)        &  11.23(24)       & 1.43(7)        & 2.493(11)            &  3.3(6)             \\  
               211392837   &   3042  &   SM    &  47.54(86) &  4.81(3)  & 4618(70)  & 10.85(21) & 1.62(10)    & 2.576(35)          & 10.33(20)       &  1.47(9)         &  10.20(28)       & 1.42(11)       & 2.571(9)             &  3.3(9)             \\  
               211413623   &   4005  &   SM    &  64.84(83) &  6.28(3)  & 4702(143) &  8.77(17) & 1.46(9)     & 2.715(74)          &  8.39(17)       &  1.33(8)         &   8.40(17)       & 1.33(8)        & 2.714(7)             &  4.2(9)             \\  
               211396385   &   1033  &   BM    &  77.4(1.2) &  7.00(4)  & 4808(290) &  8.50(29) & 1.65(17)    & 2.80(16)           &  8.23(28)       &  1.55(16)        &   8.16(17)       & 1.50(9)        & 2.792(8)             &  2.7(6)             \\  
               211414300   &   1011  &   SM    &  78.8(1.3) &  7.19(8)  & 4709(69)  &  8.12(15) & 1.52(10)    & 2.800(36)          &  7.78(14)       &  1.39(10)        &   7.79(23)       & 1.40(10)       & 2.798(7)             &  3.5(9)             \\  
               211408346   &   2006  &   SM    &  98.7(2.3) &  8.17(13) & 4723(132) &  7.90(22) & 1.80(19)    & 2.898(70)          &  7.59(21)       &  1.66(17)        &   7.42(28)       & 1.57(14)       & 2.892(10)            &  2.3(7)             \\  
               211410231   &   3011  &   SM    & 103.1(3.5) &  8.87(9)  & 4830(147) &  7.08(26) & 1.53(18)    & 2.922(86)          &  6.87(26)       &  1.44(17)        &   6.75(29)       & 1.36(16)       & 2.914(14)            &  3.7(1.7)           \\  
               211412928   &   4010  &   SM    & 117.8(1.5) &  9.74(5)  & 4817(149) &  6.70(13) & 1.57(10)    & 2.979(85)          &  6.50(13)       &  1.47(9)         &   6.55(11)       & 1.49(7)        & 2.979(6)             &  2.7(5)             \\  
               211411629   &   3004  &   BM    & 196(48)    & 14.43(13) &           &           &             &                    &                 &                  &                  &                &                      &                     \\  
               211414687   &   5010  &   SM    & 203.0(1.5) & 15.10(6)  & 4850(73)  &  4.82(5)  & 1.40(5)     & 3.217(43)          &  4.70(5)        &  1.33(5)         &   4.77(5)        & 1.38(4)        & 3.220(3)             &  3.6(4)             \\  
               211416749   &   4011  &   SM    & 234.3(1.3) & 16.76(5)  & 4851(80)  &  4.51(5)  & 1.41(5)     & 3.280(48)          &  4.40(4)        &  1.34(4)         &   4.50(5)        & 1.42(4)        & 3.282(3)             &  3.2(3)             \\  
               211421954   &   3019  &   SM    & 246.1(2.3) & 17.47(6)  & 4889(78)  &  4.38(5)  & 1.41(6)     & 3.303(48)          &  4.29(5)        &  1.35(5)         &   4.35(6)        & 1.39(5)        & 3.303(5)             &  3.4(5)             \\  
               211409560   &   4009  &   SM    & 272.2(1.7) & 19.10(7)  & 4908(111) &  4.06(5)  & 1.34(6)     & 3.347(70)          &  4.00(5)        &  1.30(5)         &   4.05(4)        & 1.34(4)        & 3.348(3)             &  3.9(4)             \\  
               211388537   &   2052  &   SM    & 287.6(8.7) & 20.15(30) & 5015(71)  &  3.90(12) & 1.32(15)    & 3.376(52)          &  3.91(12)       &  1.33(15)        &   3.88(15)       & 1.31(12)       & 3.375(12)            &  4.2(1.5)           \\  
               211403248   &   2035  &   SM    & 305.5(3.0) & 21.45(9)  & 4963(80)  &  3.64(5)  & 1.21(5)     & 3.400(54)          &  3.64(5)        &  1.21(5)         &   3.57(6)        & 1.17(5)        & 3.398(5)             &  6.5(1.1)           \\  
               211415364   &   5014  &   SM    & 463(80)    & 28.29(8)  & 4940(122) &  3.16(55) & 1.39(72)    & 3.58(12)           &  3.13(54)       &  1.35(71)        &   2.95(13)       & 1.13(14)       & 3.550(16)            &  7.4(3.6)           \\  
               211411922   &   3017  &   SM    & 559(66)    & 36.34(24) & 5158(117) &  2.36(28) & 0.96(34)    & 3.67(12)           &  2.44(29)       &  1.02(37)        &   2.57(10)       & 1.20(13)       & 3.699(15)            &  5.6(2.6)           \\  
               211409088   &   6012  &   SM    & 562(17)    & 33.02(11) & 5060(64)  &  2.85(9)  & 1.38(13)    & 3.669(51)          &  2.87(9)        &  1.40(13)        &   2.79(6)        & 1.29(7)        & 3.655(8)             &  4.3(1.0)           \\  
            ***211414203   &  10006  &   SM    & 663(308)   & 46.35(55) & 5190(77)  &  1.73(80) & 0.61(85)    & 3.75(31)           &  1.79(83)       &  0.65(91)        &   2.12(4)        & 1.10(5)        & 3.821(7)             &  7.9(1.3)           \\  

\tableline											     
\end{tabular}
\tablenotetext{}{Uncertainties are shown in compact bracket form:
  e.g. $2.35(5)=2.35\pm 0.05$, $2.35(15)=2.35\pm 0.15$, $1.297(32)=1.297\pm 0.032$, $15.6(1.3)=15.6\pm 1.3$, }
\tablenotetext{a}{See \citet{Huber16}.
  (sorted by \numax).}
\tablenotetext{b}{See \citet{Geller15}; includes cross ID to \citet{Sanders77}.} 
\tablenotetext{c}{Classification and membership from
  radial velocity \citep{Geller15}; SM: single member; BM:
  binary member.} 
\tablenotetext{*}{Red clump star according to CMD.}
\tablenotetext{**}{Marginal detection.}
\end{center}}
\end{table*}

\subsection{Seismic radius, mass, and age}
First, we use the (uncorrected) asteroseismic scaling relations,
\dnu\ $\propto M^{0.5}/R^{1.5}$ and \numax\ $\propto
M/(R^2 T^{0.5}_{\mathrm{eff}})$ to determine radius, mass, and $\log g$ for each star
(Table~\ref{tab1}, columns 7-9).  
Because the \dnu-scaling is known to show a
temperature and metallicity-dependent bias based on stellar models
\citep{White11,Miglio13,Sharma16,Guggenberger16} we apply the required correction using 
%public correction software \citep{Sharma16}; this has
the public correction software by \citet[][ corresponding to option 3 in
their Table 1]{Sharma16}\footnote{Further details about, and access
to, the \dnu-correction source code and grid can be found at
http://www.physics.usyd.edu.au/k2gap/Asfgrid/}; this has 
been shown to bring seismic masses (and radii) in better agreement with
independent determinations from eclipsing 
binaries and interferometry \citep[compare option 1 and 3 of Table 1 in~][~for an
overview]{Sharma16}.  The re-derived (corrected) scaling
relation-based radii and masses are listed in Table~\ref{tab1} (columns 10-11). 
We investigate \teff-related systematics by adopting the SDSS-DR12 \teff\
values, which on average are $90\,$K hotter than our main IRFM-based
values. This shifts the radius ($\sim$ 1\%), mass (2-3\%), and  
$\log g$ ($\sim$ 0.003$\,$dex) to larger values, of all scaling relation-based results.

\begin{figure}
\includegraphics[width=8.8cm]{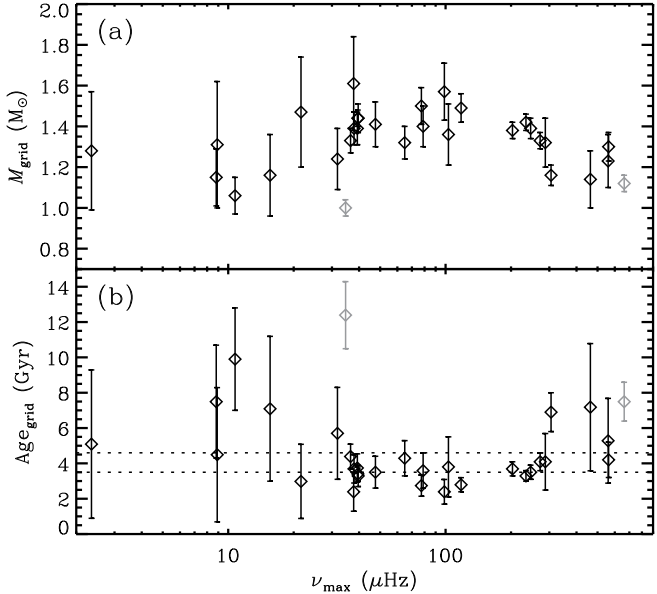}
\caption{(a): $M_\mathrm{grid}$ versus \numax\ including $1\sigma$
  errorbars. Gray symbols show stars not included in the calculations of
  the cluster's average properties (Section~\ref{averages}).
  (b): As (a) but for Age$_\mathrm{grid}$.  Horizontal dotted lines bracket
  the typical cluster-averaged age range quoted in the literature (Table~\ref{tab2}).
\label{mass-age_vs_numax}} 
\end{figure}

In order to also obtain ages we need to apply stellar models, which 
also provide more precise, but model-dependent, determinations of radius, mass, and $\log g$. 
%(REFS, CHAPLIN14,PINSONNEAULT14,SILVA12,SILVA15 DONT SEEM TO MAKE DIRECT
%COMPARISON BETWEEN SCALING AND GRID-BASED PRECISIONS FOR GIANTS...IDEAS
%FOR REFERENCES?). 
For this we use the BAyesian STellar Algorithm 
\citep[BASTA,][]{Silva15} with a grid of BaSTI isochrones that
includes both RGB and RC models \citep{Pietrinferni04}. To avoid biases
arising from the \dnu-scaling we correct the \dnu\ values in our BaSTI
models applying the prescription of Serenelli (in preparation), which
is based on computed oscillation frequencies 
%at different metallicities and evolutionary phases 
(analogous to the approach by \citet{Sharma16}). For
each star we adopt the average composition from 
\citet{Pace08} ([Fe/H$]=0.03\pm 0.04\,$dex). %DR12 ([Fe/H$]=0.08\pm 0.03\,$dex).
The results are listed in Table~\ref{tab1} (columns 12-15) and we show the mass
and age as a function of \numax\ in Figure~\ref{mass-age_vs_numax}.
Here the \teff-related systematics are negligible compared to the quoted
uncertainties except for a slightly lower SDSS-DR12-based age ($\lesssim
0.08\,$Gyr). %radius ($<0.5\,$%), mass ($<0.005\,$\msol), $\log g$ ($<0.001\,$dex) 
To highlight metalicity-related systematics we repeat the gridmodelling
with [Fe/H$]=0.08\pm0.03\,$dex (the average from SDSS-DR12), which also show
negligible shifts, even for age ($+0.03\,$Gyr). 
%radius ($<0.5$\%), mass ($<0.005$\msol), logg ($<0.001$dex)  

\subsection{Average cluster properties}\label{averages}
In the following we present weighted average cluster properties and
uncertainties on their mean, and accounting for the systematics described above.
To calculate cluster averages we ignore the results from star 211414203
because it is a marginal detection, and 211406541 because its
evolutionary state is ambiguous; it appears as an RGB star in the CMD but
as an RC star in a $\log g_\mathrm{seismic}$-\teff\ diagram
(Figure~\ref{kiel_dnu-numax_amp-numax}b). 
There is an indication in the power spectrum that the \numax\
measurement could be underestimated, which we believe is the most likely
cause of its deviant mass, and hence age (Table~\ref{tab1}). %, perhaps related to its binarity.  
We note that 211406540 appears marginally discrepant
(Figure~\ref{kiel_dnu-numax_amp-numax}a, RC close-up), but do not exclude it.

From the seismically-determined radii, we calculate the cluster's
distance modulus, $(m-M)_0$, to be $9.61\pm 0.03\,$mag ($830\pm 11\,$pc), $9.57\pm
0.03\,$mag ($816\pm 11\,$pc), and $9.55\pm 0.03\,$mag ($811\pm 11\,$pc)
based on $R_\mathrm{sc}$, $R_\mathrm{corr}$, and $R_\mathrm{grid}$,
respectively (Table~\ref{tab1}).  
From  $R_\mathrm{corr}$ we see good agreement between the distances based
on RGB ($9.57\pm 0.03\,$mag), and RC ($9.59\pm 0.07\,$mag) stars. 
Bolometric corrections were performed
using the calibrations of \citet{BessellWood84} and \citet{Flower96}, taking
$\sigma(T_\mathrm{eff})$ into account, and we assumed $A_V=3.1E(B-V)$ and
neglected the uncertainty in apparent magnitude.  
In comparison, literature values fall typically in the range
%9.47-9.72 (785-880pc)
9.5-9.7 mag (795-$870\,$pc) \citep[see][~and references therein]{Geller15}.
%$m-M=9.62$ \citep{VandenBergStetson04} to $m-M=9.70\pm0.05$ \citep{Sarajedini09}. 
%VanderBerg14':9.69
%VanderBerg08:9.70
%Geller15 summarieses 800-900pc range in lit ~(9.52-9.77 mag)
%THE ACTUAL REFS IN GELLER15 AND THEIR VALUES
%m-M_o=9.48Janes85
%m-M_o=9.61 Nissen87
%m-M_o=9.72 MMJ93
% m-M_V=9.85 MMJ93
%   m-M_?=9.6 Carraro94
%m-M_o=9.47 Fan96
%m-M_o=9.60 Sanquist04
% m-M_V=9.72 Sadquist04
%   m-M_?=9.7 Balaguer07
%   m-M_?=9.63 Pasquini08
%m-M_0=9.70 Sarajedini09
Adopting SDSS-DR12 \teff\ values increases our distance 
by $0.10\,$mag ($35\,$pc), while changing [Fe/H] has a negligible
effect.  Changing reddening by 0.01 \citep{Taylor07} will change distance
by $0.03\,$mag.  

We find the average seismic mass of stars below the RGB bump to
be $M_\mathrm{sc}=1.39\pm 0.02\,$\msol, $M_\mathrm{corr}=1.34\pm 0.02\,$\msol, and
$M_\mathrm{grid}=1.36\pm 0.02\,$\msol.  
These are all in agreement with the expected lower-RGB mass, which we derive
by extrapolating the dynamic mass, $M_\mathrm{EB}=1.31\pm 0.05\,$\msol\
\citep{Gokay13}, of the eclipsing binary HV Cnc/S986/WOCS4007
\citep{Sanders77,Geller15} located near the cluster turn-off ($V=12.73$,
$B-V=0.55$; Figure~\ref{cmd}).   
To extrapolate, we add the mass difference between the location of the
eclipsing binary and that of the RGB along the BaSTI isochrone, which is
%V=12.73 in WOCS
%BV=0.55 in WOCS
%CMD shows the mass of iso in vicinity of S986 is 1.27-1.29Msol 
%and for RGB bottom it is 1.35 and for RGB tip is is 1.38
0.05-0.09\msol\ for the RGB below the bump, hence resulting in expected
masses of about 1.36-1.40\msol.  Because the uncertainty of the 
expected mass is at least 0.05\msol (from $\sigma(M_\mathrm{EB})$, which is
probably underestimated), we are not able
to make strong conclusions on which seismic method provides the most
accurate mass. 
%and 0.09-0.11\msol\ for the RGB tip/RC, resulting in 1.40-1.42\msol.

The average RC star mass is $M_\mathrm{sc}=1.37\pm 0.09\,$\msol, 
$M_\mathrm{corr}=1.40\pm 0.09\,$\msol, and $M_\mathrm{grid}=1.40\pm 0.03\,$\msol,
which for the latter two is 0.04-0.06\msol\ more massive than for the lower
RGB. The RC-RGB mass difference is expected to be only 0.02-0.05\msol,
assuming no mass loss along the isochrone. 
Hence, we see no evidence of significant mass loss along the RGB. 
This seems to align with the seismic-based results of
\citet{Miglio12} who concluded no or little mass loss (equivalent to
Reimers $\eta$ below 0.2) for the open clusters NGC~6819 and NGC~6791,
which bracket the age of M67.
%With eta=0.2 => RC_noLoss-RC_loss = 0.07-0.08\msol.
%It is evident from Figure~\ref{kiel_dnu-numax_amp-numax}a (RC close-up)
%that the data can not distinguish between no mass loss (grey curve) and the
%$\eta=0.2$ case (red curve).

Finally, we obtain an average age of Age$_\mathrm{grid}=3.46\pm
0.13\,$Gyr for our giants. This is on the lower side compared to 
traditional
isochrone fitting results ranging 3.6-4.6$\,$Gyr \citep{VandenBergStetson04},
3.5-4.0$\,$Gyr \citep{Sarajedini09}, chromospheric activity-based ages
3.8-4.3$\,$Gyr \citep{Barnes16}, and recent K2-based gyrochronology results of
$3.7\pm0.3\,$Gyr \citep{Gonzalez16} and $4.2\pm 0.2\,$Gyr
\citep{Barnes16}, but individual star ages are statistically compatible with those from
other methods (Table~\ref{tab1}, Figure~\ref{mass-age_vs_numax}b).  
We note that the small uncertainty in the adopted average
metallicity tends to favor a grid-modeling solution in a single 
metallicity value, potentially biasing our result. %, especially for the age. 
Also, model-dependent age systematics are not taken into
account, making our uncertainties underestimated.
We would caution adopting this age given the poor match of the $3.5\,$Gyr
isochrone at the turn off in Figure~\ref{cmd}, which we attribute 
%INCLUDE NEXT TWO \bf ADDITIONS AFTER ACCEPTANCE 
%{\bf
partly to uncertainties in the convective core overshoot \citep{Dinescu95},
and partly
%}
to the lack of 
%{\bf 
stronger age contraints from
%}
turn-off stars with seismic measurements, as concluded for the Hyades (Lund
et al. in preparation). 
We summarize the results of this section in Table~\ref{tab2}. 
%\begin{table*}
%{\footnotesize
%\begin{center}
%\caption{Average cluster properties. \label{tab2}}
%\begin{tabular}{rcccccccccccc}
%\tableline\tableline
%\multicolumn{1}{c}{$(m-M)_{0,sc}$}    & $(m-M)_{0,corr}$  & $(m-M)_{0,grid}$    & Litt. range$^a$ & $M_\mathrm{BelowBump,sc}$ & $M_\mathrm{BelowBump,corr}$ & $M_\mathrm{BelowBump,grid}$ & $M_\mathrm{EB,extrapol}$ & $M_\mathrm{RC,sc}$ & $M_\mathrm{RC,corr}$ & $M_\mathrm{RC,grid}$ & Age$_\mathrm{grid}$ & Litt. range$^b$  \\
%%\multicolumn{1}{c}{ID}      & (\muhz) & (\muhz) & (IRFM)  & (\rsol)       & (\msol)        & (cgs)              & (\rsol)         & (\msol)          & (\rsol)         & (\msol)        & (cgs)                & (Gyr)               \\
%\tableline		  	  	    		                       
%               9.61(3)   &   9.57(3)  &   9.55(3)    &   9.5-9.7 &  1.39(2)      &  1.34(2)      &  1.36(2)    &   1.36(5)-1.40(5)          &     1.37(9)          &   1.40(9)        &   1.40(3)        &    3.46(13)        &   3.5-4.6             \\   
%\tableline											     
%\end{tabular}
%\tablenotetext{a}{See review in \citet{Geller15}.}
%\tablenotetext{b}{\citet{VandenBergStetson04},\citet{Sarajedini09},\citet{Barnes16},\citet{Gonzales16}.}
%\end{center}}
%\end{table*}
\begin{table}
{\footnotesize
\begin{center}
\caption{Average cluster properties. \label{tab2}}
\begin{tabular}{rcccccccccccc}
\tableline\tableline
                            &  This work & Literature     \\
\tableline		  	  	    		                       
$(m-M)_\mathrm{0,sc}$/mag     &   9.61(3) &                 \\
$(m-M)_\mathrm{0,corr}$/mag   &   9.57(3) & 9.5-9.7$^a$     \\
$(m-M)_\mathrm{0,grid}$/mag   &   9.55(3) &                 \\
                            &           &                 \\
$M_\mathrm{RGB,sc}$/\msol     &   1.39(2) &                 \\
$M_\mathrm{RGB,corr}$/\msol   &   1.34(2) & 1.36-1.40$^b$   \\
$M_\mathrm{RGB,grid}$/\msol   &   1.36(2) &                 \\
                            &           &                 \\
$M_\mathrm{RC,sc}$/\msol      &   1.37(9) &                 \\
$M_\mathrm{RC,corr}$/\msol    &   1.40(9) & 1.40-1.42$^b$   \\
$M_\mathrm{RC,grid}$/\msol    &   1.40(3) &                 \\
                            &           &                 \\
Age$_\mathrm{grid}$/Gyr       &   3.46(13)& 3.5-4.6$^c$     \\
\tableline											     
\end{tabular}
\tablenotetext{a}{See review in \citet{Geller15}.}
\tablenotetext{b}{Extrapolated from eclipsing binary (assuming no mass loss).}
\tablenotetext{c}{\citet{VandenBergStetson04,Sarajedini09,Barnes16,Gonzalez16}.}
\end{center}}
\end{table}

%MY NOTE
%There might be a tendency of an increasing mass from the bottom to about
%the bump on the RGB followed by a decline towards the RGB tip. Unknown
%systematic uncertainties in the scaling relations (for \numax\ in
%particular) and the temperature scale could quite possible be a significant
%contributor to this mass variation along the RGB.  Those systemetics needs
%to be fully understood and quantified before one can attempt to directly
%compare observed and modelles mass trends along the RGB including those
%arising from mass loss.  However, we do not see evidence for significant
%mass loss when comparing the inferred masses of RGB to RC stars. 

\section{Conclusion}
Our analysis of K2 campaign-5 data demonstrates clear
detection of oscillations in the red giants of M67,
and confirms previous claims of tentative detections in a few bright giants.
The high quality of the K2 data enables us to 
measure global asteroseismic properties of stars in M67 for the
first time.  From these we infer the stellar radius (hence distance),
mass, and age.  The distance and RGB mass are in agreement with literature
values based on isochrone-fitting and the dynamical mass of a near turn-off (early subgiant) star
in an eclipsing binary system.  The seismic-informed age is on the lower
end of independent determinations; %, such as
%gyrochronology and chromosheric activity as well as 
%classic isochrone fitting; 
reflecting that our seismic sample does not include
turn-off %or subgiant 
stars.
Our results lend support for the asteroseismic
scaling relations (when corrected for well-understood offsets) as ways to obtain
fundamental stellar properties. However, a more precise independent
determination of stellar mass at the 0.01-0.02\msol\ level, for example
from eclipsing binaries, would be desirable in future to conclude which seismic
approach is the most favorable.  It would also be interesting to compare
our results, distance in particular, with what will be obtained from Gaia. 

\acknowledgments
We acknowledge
Susan Agrain, Tim Bedding, Karsten Brogaard, Hans Kjeldsen, Daniel Huber, Marc
Pionsonneault, and Jie Yu for fruitful discussions and helpful comments.
Funding for the Stellar Astrophysics Centre is provided by The Danish
National Research Foundation (Grant DNRF106). The research was supported by
the ASTERISK project (ASTERoseismic Investigations with SONG and Kepler)
funded by the European Research Council (Grant agreement no.: 267864). 
D.S. acknowledges support from the Australian Research Council.
A.V. is supported by the NSF Graduate Research Fellowship, Grant No. DGE 1144152.
VSA acknowledges support from VILLUM FONDEN (research grant 10118).

\end{document}